\documentclass[times,twocolumn,final]{elsarticle}
\usepackage{framed,multirow}
\usepackage{amssymb}
\usepackage{latexsym}
\usepackage{url}
\usepackage[dvipsnames]{xcolor}
\usepackage[colorlinks]{hyperref}
\usepackage{bbding}
\usepackage{soul}
\usepackage{bibunits}
\usepackage{bm}
\usepackage{natbib}
\usepackage{array}
\usepackage{subfigure}
\usepackage{pdflscape}
\usepackage{makecell}
\usepackage{adjustbox}
\usepackage{framed,multirow}
\usepackage{threeparttable}
\usepackage{amssymb}
\usepackage{pifont}
\usepackage{algorithm}
\usepackage{layouts}
\usepackage{listings}
\usepackage{colortbl}
\usepackage{booktabs}
\usepackage{amsmath}
\usepackage{pythonhighlight}
\usepackage{fancyhdr}
\usepackage{pifont}
\makeatletter
\newcommand\footnoteref[1]{\protected@xdef\@thefnmark{\ref{#1}}\@footnotemark}
\makeatother

\newcolumntype{P}[1]{>{\centering\arraybackslash}p{#1}}
\newlength\savewidth

\def\arrvline{\hfil\kern\arraycolsep\vline\kern-\arraycolsep\hfilneg}
\definecolor{Highlight}{HTML}{39b54a}

\definecolor{iblue}{rgb}{0.06, 0.75, 1.0}
\definecolor{igray}{rgb}{0.00, 0.00, 0.00}

\definecolor{significant}{RGB}{239,134,131}

\definecolor{newcolor}{rgb}{.8,.349,.1}

\journal{Knowledge-Based Systems}

\begin{document}
	
	\begin{frontmatter}
 
		\title{NuSegDG: Integration of Heterogeneous Space and Gaussian Kernel for Domain-Generalized Nuclei Segmentation}
		
		\author[1]{Zhenye~Lou$^{\dag,}$}
		\author[2]{Qing~Xu$^{\dag,}$}
		\author[3]{Zekun~Jiang}
		\author[2]{Xiangjian~He\corref{cor1}}
        \ead{sean.he@nottingham.edu.cn}
		\author[6]{Chenxin~Li}
		\author[4]{Zhen~Chen}
		\author[5]{Yi~Wang}
		\author[7]{Maggie M.~He}
		\author[8]{Wenting~Duan}
		
		\cortext[cor1]{Corresponding author. \dag{Equal contribution.}}
		
		\address[1]{Sichuan University Pittsburgh Institute, Sichuan University, Chengdu, China}
		\address[2]{School of Computer Science, University of Nottingham Ningbo China, Ningbo, Zhejiang, China}
		\address[3]{West China Biomedical Big Data Center, West China Hospital, Sichuan University, Chengdu, China}
		\address[4]{Centre for Artificial Intelligence and Robotics (CAIR), Hong Kong Institute of Science \& Innovation, Chinese Academy of Sciences, Hong Kong SAR}
		\address[5]{School of Software, Dalian University of Technology, Dalian 116600, China}
		\address[6]{Department of Electronic Engineering, The Chinese University of Hong Kong, Hong Kong 999077, SAR, China}
		\address[7]{Department of Cardiology, Gold Coast University Hospital, QLD, Australia}
		\address[8]{School of Computer Science, University of Lincoln, Lincoln  LN6 7TS, UK}

		\begin{abstract}
Domain-generalized nuclei segmentation refers to the generalizability of models to unseen domains based on knowledge learned from source domains and is challenged by various image conditions, cell types, and stain strategies. Recently, the Segment Anything Model (SAM) has made great success in universal image segmentation by interactive prompt modes (e.g., point and box). Despite its strengths, the original SAM presents limited adaptation to medical images. Moreover, SAM requires providing manual bounding box prompts for each object to produce satisfactory segmentation masks, so it is laborious in nuclei segmentation scenarios. To address these limitations, we propose a domain-generalizable framework for nuclei image segmentation, abbreviated to NuSegDG. Specifically, we first devise a Heterogeneous Space Adapter (HS-Adapter) to learn multi-dimensional feature representations of different nuclei domains by injecting a small number of trainable parameters into the image encoder of SAM. To alleviate the labor-intensive requirement of manual prompts, we introduce a Gaussian-Kernel Prompt Encoder (GKP-Encoder) to generate density maps driven by a single point, which guides segmentation predictions by mixing position prompts and semantic prompts. Furthermore, we present a Two-Stage Mask Decoder (TSM-Decoder) to effectively convert semantic masks to instance maps without the manual demand for morphological shape refinement. Based on our experimental evaluations, the proposed NuSegDG demonstrates state-of-the-art performance in nuclei semantic and instance segmentation, exhibiting superior domain generalization capabilities. The source code is available at \url{ https://github.com/xq141839/NuSegDG}. 
\end{abstract}	
	\end{frontmatter}
 
	\section{Introduction}\label{sec:introduction}

Nuclei images are commonly obtained by various imaging modalities, including histopathology slides, fluorescence microscopy, and cryo-electron microscopy. The segmentation task based on such images is crucial for disease diagnosis and treatment planning~\cite{iqbal2023ldmres, xie2022dmcgnet}. In particular, semantic segmentation can be used to calculate the disease area. Instance segmentation aims to identify each nuclear as a separate entity within an image, allowing detailed morphological studies and advanced cellular analysis, such as cell counting. However, the inherent heterogeneity of different modalities, intricate tissue structures and tight cell clustering pose challenges in building a universal nuclei segmentation framework \cite{caicedo2019nucleus, kumar2017dataset, naylor2018segmentation, mahbod2021cryonuseg}. 

Traditional U-shape architectures adopt Convolutional Neural Network (CNN) for feature extraction and combine the predicted nuclear proxy maps with morphological post-processing methods to generate instance maps from the semantic segmentation masks \cite{graham2019hover, chen2023cpp, horst2024cellvit}. Despite these task-specific models displaying acceptable performance on the seen data, they are difficult to handle unseen domains, especially for the nuclei with different shapes and stain environments. This is because morphological operations are sensitive to the intensity distribution, unexpected artifacts, and noise. This highlights that the generalized nuclei segmentation methods should reduce the dependence on classical image processing algorithms.

The recent emergence of the Segment Anything Model (SAM) \cite{kirillov2023segment} has revolutionized segmentation tasks, offering versatile capabilities that surpass traditional methods. SAM has demonstrated exceptional generalization performance in natural image segmentation, showcasing robustness and adaptability across various scenarios \cite{zhang2024uv}. Based on this success, SAM has been applied to a range of medical imaging tasks, revealing its potential to handle diverse and complex segmentation challenges in the medical field, including organ and tissue segmentation and detecting various pathological conditions. These advancements underscore that SAM is promising to provide a robust and generalized solution for diverse medical image segmentation tasks \cite{shui2023unleashing, cheng2023sam, ma2024segment, huang2024segment}. Despite these advantages, globally fine-tuning SAM requires a large number of pixel-level annotated labels, so it is expensive and impractical for medical scenarios, especially for the specific disease or segmentation task.

Furthermore, SAM mainly adopts interactive prompt modes (e.g., point and box) to guide the segmentation decoding. Although the box mode enables SAM to provide accurate segmentation masks, it is sensitive to the precision of manual annotations and is labor-intensive in nuclei segmentation tasks as each nuclei image usually contains hundreds of cells. On the other hand, the point model is labor-saving, which asks users to click the desired segmentation area. However, current studies have proven that using only one positive point of every cell as the prompt is difficult to drive SAM predicting satisfactory segmentation masks \cite{zhang2024segment, meng2024nusea}. Therefore, the point prompt mode should be further optimized in nuclei segmentation tasks. 

To address these limitations in nuclei image segmentation, we propose a domain-generalizable framework for semantic segmentation and automatic instance map conversion, abbreviated to NuSegDG. It is comprised of three modules: a Heterogeneous Space Adapter (HS-Adapter), a Gaussian-Kernel Prompt Encoder (GKP-Encoder) and a Two-Stage Mask Decoder (TSM-Decoder). Specifically, HS-Adapter is used to adapt SAM from natural to different nuclei images and provides domain-specific feature representations by heterogeneous space integration. GKP-Encoder utilizes a single-point prompt to generate the density map with sufficient semantic information for guiding segmentation predictions. TSM-Decoder is responsible for predicting precise semantic segmentation masks and converting them to instance maps without manual morphological image processing. \textcolor{black}{To the best of our knowledge, we are the first attempting to discover the impact of heterogeneous space integration in domain-generalized nuclei image segmentation during the fine-tuning stage. Secondly, we innovatively leverage Gaussian kernel transforms single-point annotations into high-quality density maps, providing rich semantic and positional prompts. Finally, compared to current state-of-the-art methods \cite{stringer2021cellpose, chen2023cpp, horst2024cellvit}, we first adopt a novel semantic-to-instance sequence decoding paradigm to generate final instance segmentation maps, significantly reduce the prediction complexity, and improve generalization capabilities.} The contributions of this work are summarized as follows:
\textcolor{black}{
\begin{itemize}
    \item We introduce the HS-Adapter to seamlessly harmonize knowledge transfer between natural and nuclei images and adaptively adjust the feature representation based on different nuclei domains by leveraging heterogeneous space projection. 
    \item We devise the GKP-Encoder that utilizes the labor-saving single-point prompt and Gaussian kernel to produce a density map with sufficient semantic prompt information for guiding segmentation predictions. 
    \item We design the TSM-Decoder that provides a novel semantic-to-instance sequence decoding paradigm to eliminate manual morphological shape refinement and achieve the prediction of accurate instance segmentation maps. 
    \item Our NuSegDG framework integrates HS-Adapter, GKP-Encoder, and TSM-Decoder. We conduct extensive experiments on diverse nuclei image datasets, demonstrating that NuSegDG outperforms classical nuclei segmentation methods and state-of-the-art medical SAM variants and revealing superior domain generalization capabilities.
\end{itemize}
}

\begin{figure*}[t]
  \centering
  \includegraphics[width=0.95\linewidth]{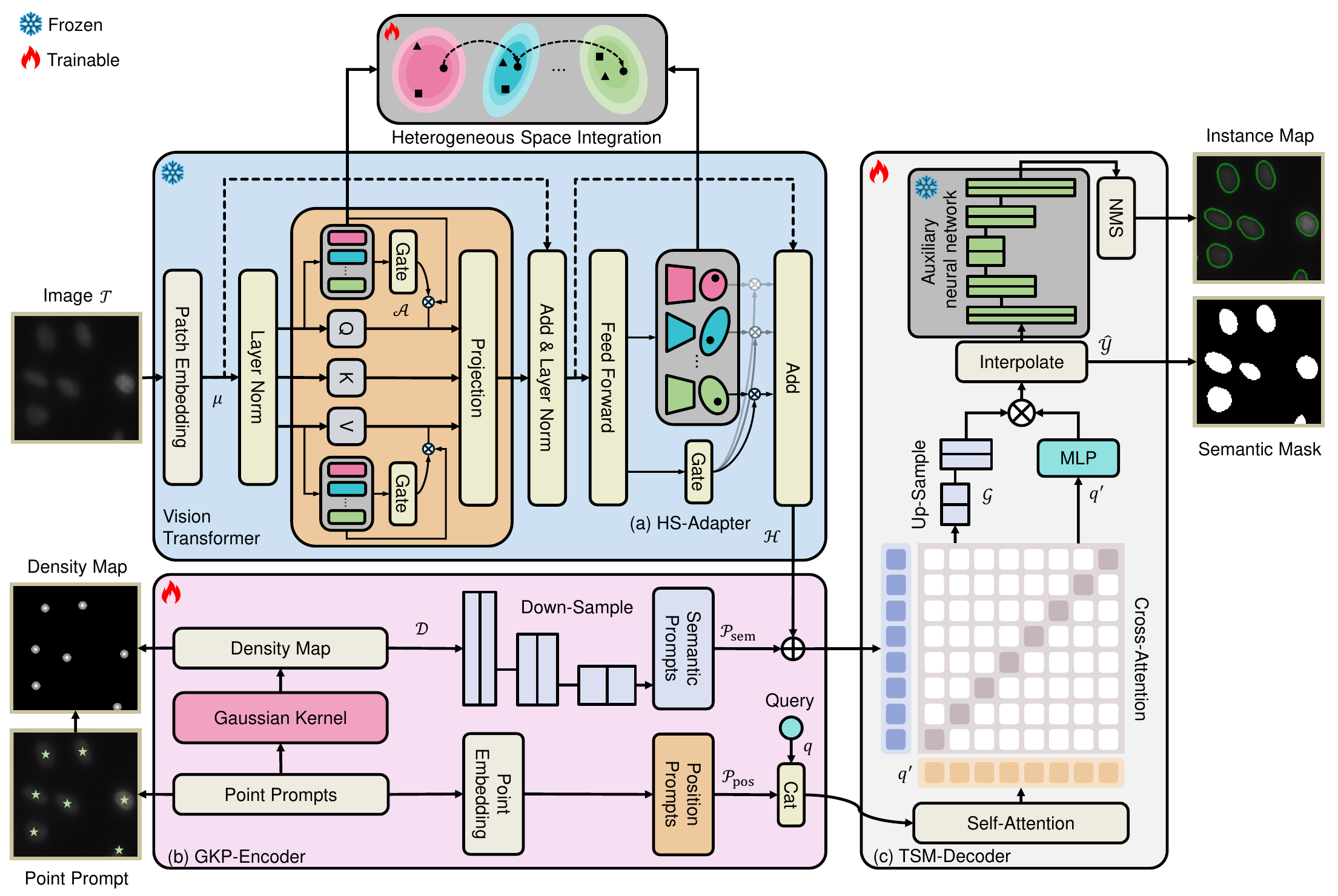}
  \caption{The overview of our NuSegDG for domain-generalized nuclei image segmentation. (a) Heterogeneous Space Adapter. (b) Gaussian-Kernel Prompt Encoder. (c) Two-Stage Mask Decoder.}
  \label{fig:csam}
  \vspace{-1.0em}
\end{figure*}

\section{Related Work}
In this section, we review the state-of-the-art nuclei segmentation architectures. Moreover, the traditional DG frameworks and recent medical foundation models are mentioned.
\subsection{Nuclei Image Segmentation}
The segmentation of nuclei in histopathology images plays an essential role in pathological analysis, enabling pathologists to make precise diagnoses \cite{chen2024sam}. It can be mainly divided into nuclei semantic segmentation and nuclei instance segmentation. The semantic segmentation focuses on the accuracy of pixel-level classification in each nuclei image, where U-Net \cite{ronneberger2015u} has made great success in this task. Over the last decade, researchers mainly focused on improving its ability of feature extraction. Early CNN series leverages the advantages of inductive bias to provide sufficient prior knowledge for accelerating model convergence \cite{ xu2023dcsau}. Vision Transformer (ViT) \cite{dosovitskiy2021an} further increases the model capacity by utilizing a self-attention mechanism to capture long-range dependencies \cite{lin2022ds, li2023lvit, chen2024transunet}. The recent Mamba-based frameworks adopted State Space Model (SSM) to optimize the computation complexity of global context \cite{ma2024u}.

In addition, the instance segmentation task aims to identify each nucleus as a distinct entity. Existing methods usually predict different types of nuclear proxy maps to synthesize instance maps. HoVer-Net \cite{graham2019hover}, Cellpose \cite{ stringer2021cellpose} and CellViT \cite{horst2024cellvit}, for instance, employed horizontal and vertical distance maps to accurately delineate the boundaries of individual nuclear in histopathology images. PROnet \cite{nam2023pronet} utilized offset maps to enhance the delineation of nuclei boundaries. TSFD-Net \cite{ilyas2022tsfd} and CPP-Net \cite{chen2023cpp} additionally used boundary maps as auxiliary supervisions. Despite their advancements, these methods often require complex post-processing, such as manual morphology operations and thresholding algorithms, to manually synthesize instance maps, so they hinder the generalization capability of models to unseen domains \cite{shui2023unleashing}. Our proposed NuSegDG framework addresses these limitations by converting fundamental semantic segmentation masks to instance maps automatically, thereby demonstrating outstanding domain generalization performance across diverse nuclei image domains.

\subsection{SAM for Generalized Medical Image Segmentation}
The generalizability of neural networks is crucial for medical image segmentation~\cite{li2022domain}. Existing methods mainly utilize multi-source domain adaptation \cite{hu2022domain} and federal learning \cite{liu2021feddg} to address the Domain Generalization (DG) problem. The recent Segment Anything Model (SAM) \cite{kirillov2023segment} is a novel interactive architecture that leverages both sparse prompts (e.g., point, box and text) and dense prompts (e.g., mask) to guide the prediction of segmentation masks. Due to its large image encoder, SAM demonstrates robust feature extraction capabilities, enabling outstanding zero-shot generalization across diverse natural image segmentation tasks. On this basis, current studies have explored the potential of SAM in medical image segmentation tasks. For instance, MedSAM \cite{ma2024segment} and SAMMI \cite{huang2024segment} globally fine-tuned SAM on more than 10 medical visual modality datasets, achieving notable generalization capabilities with bounding box prompts. However, globally fine-tuning SAM is computationally intensive and needs sufficient training samples due to its large ViT encoder, which is not efficient for nuclei image segmentation tasks. To address this issue, Parameter-Efficient Fine-Tuning (PEFT) techniques have received the most attention from researchers. Methods such as Low-Rank Adaptation (LoRA) \cite{hu2021lora} and Conv-LoRA \cite{zhong2024convolution} injected a set of trainable low-rank matrices into the attention layer of ViT to update the feature representation. Adapter \cite{houlsby2019parameter} is another common approach used to fine-tune the foundation model \cite{wu2023medical, lin2023samus}. Although these methods reveal their power in homogeneous domain generalization tasks, nuclei images in different domains have disjoint label spaces. Our approach utilizes heterogeneous space mapping to harmonize the feature representation of SAM between natural and nuclei images.

Moreover, various SAM models \cite{ma2024segment, zhang2024segment} have demonstrated the necessity of using bounding boxes as prompts to achieve optimal segmentation results in medical imaging. Conversely, relying on single-point prompts often fails to provide sufficient contextual information for accurate segmentation, especially in complex and dense nuclei images \cite{huang2024segment, meng2024nusea}. To overcome the limitations of single-point prompts, existing studies introduced extra units, such as YOLO-NAS \cite{terven2023comprehensive} and GroundingDino \cite{liu2023grounding}, to generate prompts. They perform object detection to identify points or bounding boxes, which are then used as prompts for SAM. However, due to the heterogeneity of nuclei images across different datasets, these single-task models often struggle to provide correct prompts, leading to sub-optimal segmentation results. On the contrary, our NuSegDG uses labor-saving single-point annotation to generate sufficient position and semantic prompts, enhancing the generalization capability.

\section{Method}
\subsection{Overview of NuSegDG}
In DG, $\mathcal{S} = \{ \mathcal{S}_{k} = \{ (\mathcal{X}_k, \mathcal{Y}_k) \}, k=1, 2,\cdots, K\}%_{k=1}^{K}
$ is denoted as the set of $K$ distinct source domains, where $\mathcal{X}_k$ is the image in the $k$-th
source domain and $\mathcal{Y}_k$ is the segmentation mask of $\mathcal{X}_k$. Let $\mathcal{X}=\{\mathcal{X}_k\}_{k=1}^K$ and $\mathcal{Y}=\{\mathcal{Y}_k\}_{k=1}^K$. The goal of DG is to train a model $f_\theta: \mathcal{X} \rightarrow \mathcal{Y}$, where $\theta$ represents learned parameters. The trained model can be generalized to an unseen target domain $\mathcal{T}$ with high performance.

As illustrated in Fig. \ref{fig:csam}, we present the overview of NuSegDG for domain-generalized nuclei image segmentation. Given a nuclei image from the $k$-th domain, we first utilize the Heterogeneous Space Adapter (HS-Adapter) to update the attention computation and feature representation of SAM. The generated image embeddings are then delivered to the Gaussian-Kernel Prompt Encoder (GKP-Encoder) that adopts the single-point annotation to generate sufficient position and semantic information prompts for guiding segmentation decoding. Following this, the Two-Stage Mask Decoder (TSM-Decoder) leverages these prompts and image embeddings to produce a precise semantic segmentation mask and then automatically converts them to an instance map without the demand for laborious manual post-processing operations.

\subsection{Heterogeneous Space Adapter} \label{m1}
Recent studies \cite{zhang2024uv, huang2024segment} have highlighted the impressive generalized segmentation capabilities of SAM \cite{kirillov2023segment}, facilitated by its large-scale image encoder. Especially, the conventional Adapter \cite{houlsby2019parameter} and LoRA \cite{hu2021lora} have been widely used to adapt SAM to medical image segmentation \cite{wu2023medical, zhang2023customized}. However, such homogeneous space mapping methods are difficult to learn heterogeneous relationships~\cite{huang2024hst} between different nuclei domains. To tackle the issue, we propose the HS-Adapter that leverages heterogeneous space integration to enhance the domain-specific feature representation of nuclei images. Specifically, the input image is first converted into a set of 2D patch embeddings $\mu \subset \mathbb{R}^{(\frac{H\times W}{n})\times d}$, where $H$ and $W$ are height and width of the image, $n=16\times 16$ and $d=768$ stand for the patch size and channels of each patch embedding, respectively. To improve the information interaction within Multi-Head Attention (MHA) layers, the HS-Adapter respectively concatenates learnable parameters $W_{\rm que} = \{(E^i_{\rm que}, \theta^i_{\rm que})\}_{i=1}^{N}$ and $W_{\rm val} = \{(E^i_{\rm val}, \theta^i_{\rm val})\}_{i=1}^{N}$ with the query $\mathcal{Q}$ and value $\mathcal{V}$ branches of SAM, where $E^i_{\rm que}$ and $E^i_{\rm val}$ are projection layers that map embeddings $\mu$ into feature spaces with $i$-th target mapping channel, $\theta^i_{\rm que}$ and $\theta^i_{\rm val}$ are up-projections. Additionally, we place the softmax operation $\delta$ on $\mu$ to calculate the weight of each feature space. Finally, $N$ weighted different feature spaces are merged into a heterogeneous space that is used to update the original query and value projection layers of SAM, guiding the computation of attention maps as: 
\begin{equation}
        \mathcal{A} = \delta(\frac{(\mathcal{Q}(\mu) \smallfrown h_{\rm que}) \cdot {\mathcal{K}(\mu) }^{T}}{\sqrt{d}})\cdot (\mathcal{V}(\mu) \smallfrown h_{\rm val}),
\end{equation}
where
\begin{equation}
     h_{\rm que} = \mathcal{Q}(\mu) + \sum_{i=1}^N \delta(\mu)_i \theta_{\rm que}^i (E_{\rm que}^i  (\mu)),
\end{equation}
\begin{equation}
     h_{\rm val} = \mathcal{V}(\mu) + \sum_{i=1}^N \delta(\mu)_i \theta_{\rm val}^i (E_{\rm val}^i  (\mu)),
\end{equation}
$\mathcal{K}$ is the key branch of SAM, $\delta(\cdot)_i$ is the $i$-th component of $\delta(\cdot)$ and $\smallfrown$ is an concatenation operation. In addition to updating the attention computation, we apply heterogeneous space integration to the feed-forward network $\mathcal{F}_{\rm ffn}$ for learning domain-specific embeddings. The final image embeddings $\mathcal{H}$ are defined by:
\begin{equation}
    \mathcal{H} = \mathcal{A} + \mathcal{F}_{\rm ffn}(\mathcal{A}) + \sum_{i=1}^N \delta(\mathcal{A})_i \theta_{\rm ffn}^i(\phi(E_{\rm ffn}^i  (\mathcal{A}))),
\end{equation}
where $\{E_{\rm ffn}^i\}_{i=1}^N$ is a set of learnable linear layers that projects $\mathcal{A}$ into the different dimensions for the construction of heterogeneous space, $\{\theta_{\rm ffn}^i\}_{i=1}^N$ is a set of the up-projections used to align the dimension with $\mathcal{A}$, and $\phi$ is the nonlinear activation function. Compared to the conventional parameter-efficient fine-tuning techniques, the HS-Adapter performs better in learning heterogeneous relationships between different nuclei domains by using multi-dimensional projection, enhancing the representation of domain-specific knowledge. Overall, our proposed HS-Adapter significantly reduces the number of parameters during the fine-tuning stage.

\subsection{Gaussian-Kernel Prompt Encoder} \label{m2}
The original SAM \cite{kirillov2023segment} and medical SAMs \cite{wu2023medical,ma2024segment,huang2024segment} mainly rely on manual box prompts to guide the model in predicting accurate segmentation masks. Despite its advantages, this prompt mode is sensitive to the localization of boxes. Minor labeling errors can significantly reduce the quality of generating segmentation masks. Therefore, the precise manual box annotation is impractical in nuclei segmentation tasks as a histopathological image usually contains thousands of nuclei and tight cell clusters in clinical scenarios. In this paper, we introduce the GKP-Encoder that leverages single-point prompts to produce a high-quality density map, providing additionally sufficient semantic information prompts to assist segmentation decoding.

% $\mathcal{D}=\{\{\mathcal{D}_{z,j}\}_{z=1}^W\}_{j=1}^H$

Given $L$ cell positions: $\{(x_l, y_l)\}_{l=1}^L$ in a nuclei image, where $x_l, y_l \in \mathbb{N}$, the corresponding density map $\mathcal{D}=\{\mathcal{D}_{z,j}\} \in \mathbb{R}^{H \times W}$ \cite{xu2023sppnet} is defined by:
\begin{equation}
    \mathcal{D}_{z,j} = \sum_{l=1}^L G_{\sigma}(z-x_l, j-y_l),
\end{equation}
where
\begin{equation}
    G_{\sigma}(z-x_l, j-y_l) = C_{\rm norm} \cdot e^{-\frac{(z-x_l)^2 + (j-y_l)^2}{2\sigma^2}},
    \label{eq6}
\end{equation}
$z\in\{0,1,\cdots, W\}$, $j\in \{0,1,\cdots,H\}$, $\sigma^2$ is the isotropic covariance, and $C_{\rm norm}$ is a normalization constant. In Eq. \ref{eq6}, 
$G_{\sigma}(\cdot)$
%\in \mathbb{R}^{(2r+1) \times (2r+1)}$ 
stands for a normalized 2D Gaussian kernel, and 
\begin{equation}
\sum_{z-x_l=-r}^{r} \sum_{j-y_l=-r}^{r} G_{\sigma}(z-x_l, j-y_l) = 1, 
\end{equation}
where $r \in \mathbb{Z}$ determines the kernel size of %\in \mathbb{R}^
${(2r+1) \times (2r+1)}$. To fit different sizes of nuclei and provide sufficient semantic information, the parameter $r$ is set to 10 in our study. In the next step, we utilize a small convolutional network to transform $\mathcal{D}$ to a set of high-quality semantic information prompt embeddings $\mathcal{P}_{\rm sem} \in \mathbb{R}^{(\frac{H \times W}{n}) \times 256}$, where 256 is the channel number. The computation is formulated as:
\begin{equation}
    \mathcal{P}_{\rm sem} = \phi(F_{\rm conv}(\phi(F_{\rm norm}(F_{\rm conv}(\mathcal{D}))))),
\end{equation}
where $F_{\rm conv}$ is a $2 \times 2$ convolutional layer with the stride 2, $F_{\rm norm}$ is LayerNorm and $\phi$ represents GELU activation function. Moreover, the provided cell positions are used to generate additional position prompt embedding $\mathcal{P}_{\rm pos}$ using the sparse prompt encoder of SAM, where $\mathcal{P}_{\rm pos} \in \mathbb{R}^{L \times 256}$ stands for the sum of a positional encoding of the location and learnable embeddings. In this way, the proposed GKP-Encoder, driven by the single-point annotation, not only is labor-saving compared to the box annotation but also provides efficient semantic prompts $\mathcal{P}_{\rm sem}$ and position prompts $\mathcal{P}_{\rm pos}$ for guiding segmentation decoding.

\subsection{Two-Stage Mask Decoder} 
In the last decade, U-shape hierarchical decoders \cite{graham2019hover, ilyas2022tsfd, horst2024cellvit} have been widely used for the prediction of nuclei semantic and instance segmentation masks. For the latter, previous methods usually utilized morphological post-processing methods to detect each cell based on the generated nuclear proxy maps. However, such operations require laboriously manual parameter adjustment when facing different nuclei
domains, degrading the generalization capabilities of models. On the other hand, current medical SAMs \cite{ma2024segment, huang2024segment, shui2023unleashing} adopted a sequential inference algorithm to recognize each target object in images, so they are time-consuming for nuclei instance segmentation tasks involving a large number of cells. To address this issue, we propose the TSM-Decoder that improves the efficiency of producing instance maps by focusing on the prediction of precise semantic segmentation masks. Specifically, we first create trainable query embeddings $q \in \mathbb{R}^{C \times 256}$ to save the decoding information. Different from SAM \cite{kirillov2023segment}, $C$ represents the number of prediction categories instead of multi-layer masks as histopathology images may include different types of nuclei. Then, we concatenate position prompts $\mathcal{P}_{pos}$ with $q$ and perform a self-attention operation as: 
\begin{equation}
    q' = \delta(\frac{\mathcal{Q}(\mathcal{P}_{pos} \smallfrown q) \cdot {\mathcal{K}(\mathcal{P}_{pos} \smallfrown q) }^{T}}{\sqrt{d}})\cdot \mathcal{V}(\mathcal{P}_{pos} \smallfrown q), 
\end{equation}
where $q' \in \mathbb{R}^{(C+L) \times 256}$ is updated query embedding. Following this, we combine the image embedding $\mathcal{H}$ with semantic information prompts $\mathcal{P}_{sem}$: $\mathcal{H}' \gets \mathcal{H} \oplus \mathcal{P}_{sem}$, where $\oplus$ stands for the element-wise addition operation. Further, we conduct cross-attention with $q'$ to generate decoding embeddings $\mathcal{G}$, by: 
\begin{equation}
    \mathcal{G} = \delta(\frac{(\mathcal{H}'+\Psi) \cdot {(q') }^{T}}{\sqrt{d}})\cdot q' + \mathcal{H}', 
\end{equation}
where $\Psi$ is positional encodings. Similar to SAM \cite{kirillov2023segment}, we iterate this operation twice for sufficient updation. Finally, we predict the semantic segmentation mask $\hat{\mathcal{Y}}_k \in \mathbb{R}^{H \times W}$ by:
\begin{equation} 
\hat{\mathcal{Y}}_k = \rho(\mathcal{F}_{\rm inter}(\mathcal{F}_{\rm trans}(\mathcal{G}) \cdot \mathcal{F}_{\rm MLP}(q'))),
\end{equation}
where $\mathcal{F}_{\rm trans}$ is a $4 \times 4$ transpose convolution for up-sampling the decoding embeddings, $\mathcal{F}_{\rm MLP}$ represents a multilayer perceptron to perform dimensional alignment, $\mathcal{F}_{\rm inter}$ is a bilinear interpolation function to recover the shape of masks and $\rho$ is the sigmoid function. During the fine-tuning stage, we apply the weighted combination of focal loss $L_{\rm focal}$ and dice loss $L_{\rm dice}$ to supervise the predicted semantic mask $\hat{\mathcal{Y}}_k$ of different domains by: 
\begin{equation} 
    L_{\rm sem} = \alpha L_{\rm dice} + \beta L_{\rm focal},
\end{equation}
where $\alpha$ and $\beta$ respectively stand for the coefficients of focal loss and dice loss. On this basis, the prediction semantic mask can provide accurate target segmentation areas, enabling simply separating each cell by using an auxiliary neural network (e.g., StarDIST). In summary, our NuSegDG framework achieves domain generalization on both nuclei semantic and instance segmentation tasks.

\begin{table*}[!t]
    \centering
    \caption{\textcolor{black}{Comparison with state-of-the-arts on nuclei semantic segmentation (Source Domain Generalization).}}
    {\scalebox{0.74}{
    \begin{tabular}{l|c|cccc|cccc|cccc|cccc}
    \toprule
     Datasets & Manual  & \multicolumn{4}{c}{$\mathcal{S}_1$} & \multicolumn{4}{|c}{$\mathcal{S}_2$} & \multicolumn{4}{|c}{$\mathcal{S}_3$} & \multicolumn{4}{|c}{$\mathcal{S}_4$} \\
    \midrule
       Methods    &  Prompt    & Dice  & mIoU  & F1   & HD   & Dice  & mIoU  & F1   & HD   & Dice  & mIoU  & F1   & HD   & Dice  & mIoU  & F1   & HD  \\
    \midrule
    U-Net \cite{ronneberger2015u}  & \multirow{12}{*}{\ding{55}}   & 90.42 & 83.22 & 91.07 & 137.11 & 71.41 & 56.17 & 72.06 & 78.13 & 72.27 & 57.38 & 72.97 & 123.13 & 81.02 & 68.52 & 81.19 & 90.01\\
    U-Net++ \cite{zhou2019unet++}        &  & 90.85 & 83.83 & 91.33 & 117.58 & 75.72 & 61.05 & 76.15 & 73.38 & 64.66 & 49.61 & 67.56 & 148.37 & 81.80 & 69.62 & 91.97 & 80.32\\
    AttUNet \cite{schlemper2019attention}  &  & 91.01 & 84.13 & 91.41 & 112.98 & 75.81 & 61.23 & 76.13 & 74.32 & 75.98 & 61.81 & 77.28 & 131.51 & 81.34 & 68.98 & 81.69 & 80.75\\
    DCSAU-Net \cite{xu2023dcsau}         &  & 91.74 & 85.15 & 92.04 & 127.16 & 75.19 & 60.38 & 75.58 & 77.22 & 78.33 & 64.61 & 78.99 & 103.61 & 80.90 & 68.42 & 81.24 & 83.89\\
    TransUNet \cite{chen2024transunet}     &  & 91.41 & 84.65 & 91.75 & 132.94 & 76.30 & 61.83 & 76.70 & 77.33 & 76.59 & 62.38 & 77.51 & 109.36 & 82.49 & 70.47 & 82.60 & 78.14\\
    ACC-UNet \cite{ibtehaz2023acc}        &  & 90.95 & 83.96 & 91.38 & 119.39 & 77.90 & 63.93 & 78.32 & 74.79 & 66.84 & 52.32 & 70.29 & 147.97 & 82.13 & 69.91 & 82.23 & 77.89\\
    nnU-Net \cite{isensee2021nnu}          &  & 90.11 & 82.45 & 90.57 & 138.83 & 80.84 & 67.91 & 81.03 & 73.04 & 84.32 & 72.91 & 84.43 & 117.51 & 81.37 & 68.73 & 81.63 & 83.01\\
    U-mamba \cite{ma2024u}               &  & 89.96 & 82.58 & 90.66 & 127.03 & 77.38 & 63.23 & 77.74 & 73.10 & 63.62 & 47.81 & 65.44 & 167.67 & 82.43 & 70.37 & 82.57 & 82.42\\
    FedDG \cite{liu2021feddg}             &    & 90.48 & 83.25 & 91.11 & 134.97 & 74.57 & 59.77 & 75.16 & 76.74 & 74.88 & 60.45 & 76.57 & 110.62 & 81.51 & 69.17 & 81.93 & 86.56\\
    DCAC \cite{hu2022domain}             &    & 91.22 & 84.35 & 91.60 & 112.82 & 74.89 & 60.11 & 75.32 & 74.31 & 63.79 & 49.75 & 68.14 & 152.78 & 81.53 & 69.19 & 81.65 & \underline{71.69}\\
    SAC \cite{na2024segment}               &          & 91.79 & 85.16 & 92.10 & 123.01 & 81.06 & 68.25 & 81.21 & 72.68 & 83.51 & 71.74 & 83.72 & 119.45 & 81.80 & 69.36 & 81.96 & 80.83\\
    H-SAM \cite{cheng2024unleashing}        &         & 92.01 & 85.53 & 92.27 & 129.34 & 81.45 & 68.81 & 81.55 & 71.99 & 83.40 & 71.60 & 83.51 & 118.60 & 81.65 & 69.20 & 82.09 & 82.24\\
    \midrule
    SAM \cite{kirillov2023segment}         &   \multirow{6}{*}{\ding{51}}       & 89.78 & 82.14 & 90.33 & 138.58 & 76.87 & 62.63 & 77.23 & 78.83 & 79.28 & 65.91 & 79.90 & 153.47 & 77.75 & 64.43 & 78.39 & 101.65\\
    Med-SA \cite{wu2023medical}            &             & 91.72 & 85.12 & 92.08 & \underline{69.42} & 81.32 & 68.61 & 81.57 & 74.59 & 83.47 & 71.89 & 83.63 & \underline{86.62} & 83.32 & 71.59 & 83.37 & 77.74\\
    SAMed \cite{zhang2023customized}         &               & 91.32 & 84.47 & 91.63 & 71.67 & 80.06 & 66.97 & 80.46 & 77.14 & 81.75 & 69.20 & 81.78 & 101.22 & 82.23 & 70.05 & 82.32 & 76.68\\
    LeSAM \cite{gu2024lesam}               &         & 91.71 & 85.08 & 92.00 & 128.00 & 79.67 & 66.34 & 79.85 & 74.36 & 82.43 & 70.15 & 82.57 & 137.49 & 80.38 & 67.56 & 80.90 & 90.91\\
    SAMUS \cite{lin2023samus}               &            & \underline{92.07} & \underline{85.67} & \underline{92.35} & 115.18 & \underline{83.26} & \underline{71.39} & \underline{83.33} & \underline{71.14} & \underline{84.67} & \underline{73.45} & \underline{84.72} & 114.98 & \underline{83.40} & \underline{71.75} & \underline{83.70} & 79.15\\
    SAM-CL \cite{zhong2024convolution}    &             & 91.81 & 85.29 & 92.12 & 89.48 & 81.17 & 68.41 & 81.34 & 73.57 & 82.95 & 70.90 & 83.02 & 135.31 & 82.51 & 70.46 & 82.86 & 90.85\\

    \midrule
    NuSegDG                              &  \ding{51}             & \textbf{93.17} & \textbf{87.46} & \textbf{93.35} & \textbf{33.18} & \textbf{86.37} & \textbf{76.06} & \textbf{86.40} & \textbf{44.44} & \textbf{88.20} & \textbf{78.93} & \textbf{87.03} & \textbf{49.69} & \textbf{84.59} & \textbf{73.44} & \textbf{84.75} & \textbf{64.61}\\
    \bottomrule
    \end{tabular}}}
    \label{tab1}
\end{table*}

\begin{table*}[!t]
    \centering
    \caption{\textcolor{black}{Comparison with state-of-the-arts on nuclei semantic segmentation (Target Domain Generalization).}}
    {\scalebox{0.74}{
    \begin{tabular}{l|c|cccc|cccc|cccc|cccc}
    \toprule
     Datasets & Manual & \multicolumn{4}{c}{$\mathcal{T}=\mathcal{S}_1$} & \multicolumn{4}{|c}{$\mathcal{T}=\mathcal{S}_2$} & \multicolumn{4}{|c}{$\mathcal{T}=\mathcal{S}_3$} & \multicolumn{4}{|c}{$\mathcal{T}=\mathcal{S}_4$}  \\
    \midrule
       Methods &  Prompt  & Dice  & mIoU & F1   & HD & Dice  & mIoU  & F1   & HD & Dice  & mIoU  & F1   & HD & Dice  & mIoU  & F1   & HD\\
    \midrule
    U-Net \cite{ronneberger2015u}       & \multirow{12}{*}{\ding{55}} & 21.88 & 15.31 & 26.52 & 355.19 & 56.01 & 41.66 & 65.81 & 77.08 & 25.73 & 16.87 & 30.73 & 399.99 & 65.49 & 49.89 & 68.52 & 99.06\\
    U-Net++ \cite{zhou2019unet++}        &  & 25.99 & 17.53 & 31.55 & 372.39 & 59.64 & 46.03 & 67.26 & 86.35 & 20.14 & 12.50 & 22.50 & 418.79 & 66.54 & 51.04 & 68.80 & 99.63\\
    AttUNet \cite{schlemper2019attention}  & & 26.66 & 18.30 & 32.42 & 372.44 & 58.79 & 45.24 & 66.79 & 92.43 & 27.70 & 17.98 & 31.74 & 375.44 & 66.13 & 50.55 & 68.46 & 101.81\\
    DCSAU-Net \cite{xu2023dcsau}         &  & 43.42 & 30.73 & 47.49 & 246.01 & 67.17 & 52.72 & 71.83 & 78.62 & 33.66 & 22.34 & 37.27 & 307.45 & 71.58 & 56.47 & 73.25 & 94.36\\
    TransUNet \cite{chen2024transunet}     &  & 64.41 & 52.85 & 70.93 & 151.67 & 77.51 & 63.54 & 78.11 & 64.64 & 73.41 & 60.26 & 76.53 & 229.66 & 74.17 & 59.43 & 75.27 & 95.42\\
    ACC-UNet \cite{ibtehaz2023acc}        &  & 29.11 & 20.27 & 35.12 & 381.36 & 65.75 & 51.39 & 70.54 & 77.79 & 32.51 & 21.29 & 35.35 & 312.20 & 69.63 & 54.24 & 70.94 & 98.92\\
    nnU-Net \cite{isensee2021nnu}          & & 23.71 & 16.34 & 28.48 & 360.33 & 62.43 & 48.37 & 69.79 & 81.52 & 31.45 & 20.75 & 34.70 & 342.90 & 68.88 & 53.50 & 70.67 & 99.56\\
    U-mamba \cite{ma2024u}               & & 12.01 & 9.05  & 14.89 & 392.97 & 51.68 & 39.28 & 62.53 & 101.20 & 19.39 & 11.89 & 21.50 & 408.82 & 58.79 & 45.24 & 66.79 & 92.43\\
    FedDG \cite{liu2021feddg}             &  & 62.79 & 49.43 & 69.99 & 266.36 & 77.03 & 62.90 & 78.01 & 70.12 & 71.19 & 57.65 & 74.47 & 242.18 & 71.85 & 56.64 & 72.83 & 99.19\\
    DCAC \cite{hu2022domain}             &  & 50.29 & 38.01 & 55.50 & 162.13 & 68.50 & 53.71 & 72.48 & 73.95 & 29.96 & 19.48 & 33.36 & 322.49 & 69.71 & 54.36 & 71.15 & 100.21\\
    SAC \cite{na2024segment}               &    & 69.11 & 58.14 & 74.60 & 245.04 & 77.80 & 63.89 & 78.52 & 66.76 & 76.09 & 63.17 & 78.49 & 212.19 & 75.32 & 60.86 & 75.74 & 90.28\\
    H-SAM \cite{cheng2024unleashing}        &     & 76.92 & 66.15 & 80.67 & 165.25 & 78.00 & 64.17 & 78.65 & 66.07 & 76.33 & 64.03 & 78.89 & 215.47 & 77.04 & 63.00 & 77.40 & 89.83\\
    \midrule
    SAM \cite{kirillov2023segment}         &  \multirow{6}{*}{\ding{51}}  & 66.59 & 55.29 & 72.37 & 252.80 & 76.84 & 62.63 & 77.38 & 66.82 & 76.48 & 63.16 & 78.34 & 144.39 & 75.56 & 61.25 & 76.14 & 75.61\\
    Med-SA \cite{wu2023medical}            &     & 76.94 & 66.21 & 80.73 & 125.24 & 79.55 & 66.25 & 80.00 & 65.19 & 78.43 & 66.26 & 81.02 & 127.02 & 80.54 & 67.66 & 80.75 & 80.75\\
    SAMed \cite{zhang2023customized}         &   & 76.79 & 65.70 & 80.27 & 137.15 & 78.86 & 65.35 & 79.51 & 66.20 & 78.48 & 65.68 & 80.06 & 137.82 & 78.77 & 65.29 & 79.08 & \underline{69.09}\\
    LeSAM \cite{gu2024lesam}               &     & 71.80 & 61.45 & 77.55 & 207.65 & 77.83 & 63.95 & 78.58 & 67.65 & 76.48 & 63.78 & 78.80 & 205.46 & 74.72 & 60.02 & 75.06 & 91.03\\
    SAMUS \cite{lin2023samus}               &     & \underline{78.51} & \underline{68.09} & \underline{82.14} & \underline{107.32} & \underline{80.25} & \underline{67.16} & \underline{80.68} & \underline{63.27} & \underline{80.91} & \underline{68.69} & \underline{82.40} & \underline{82.33} & \underline{80.72} & \underline{67.98} & \underline{80.97} & 91.39\\
    SAM-CL \cite{zhong2024convolution}    &   & 73.48 & 63.05 & 78.65 & 157.08 & 77.51 & 63.54 & 78.11 & 64.64 & 76.99 & 63.94 & 79.24 & 119.73 & 78.65 & 65.12 & 78.95 & 74.44\\

    \midrule
    NuSegDG                              &  \ding{51}   & \textbf{80.55} & \textbf{70.71} & \textbf{84.19} & \textbf{54.81} & \textbf{82.43} & \textbf{70.23} & \textbf{82.72} & \textbf{61.36} & \textbf{82.88} & \textbf{71.24} & \textbf{83.34} & \textbf{64.56} & \textbf{83.90} & \textbf{72.49} & \textbf{84.11} & \textbf{64.38} \\
    \bottomrule
    \end{tabular}}}
    \label{tab2}
    % \vspace{-2.0em}
\end{table*}

\section{Experiments}

\subsection{Datasets and Implementations}
\subsubsection{Datasets}
To validate the effectiveness of the proposed NuSegDG, we collect DSB-2018 \cite{caicedo2019nucleus}, MoNuSeg-2018 \cite{kumar2017dataset}, TNBC \cite{naylor2018segmentation} and CryoNuSeg \cite{mahbod2021cryonuseg} datasets to perform comprehensive comparisons for domain generalization. We denote these four nuclei datasets with source domains $\mathcal{S}_1$, $\mathcal{S}_2$, $\mathcal{S}_3$ and $\mathcal{S}_4$, respectively.  The details are as follows.

\textbf{\textit{DSB-2018}} \cite{caicedo2019nucleus} dataset includes 670 nuclei images captured using fluorescence microscopy, offering a range of staining methods including DAPI, Hoechst, hematoxylin and eosin. These images are annotated with nuclear masks to facilitate segmentation tasks and vary in size.

\textbf{\textit{MoNuSeg-2018}} \cite{kumar2017dataset} dataset consists of 51 stained histopathology images from various organs, including breast, liver, kidney, prostate, bladder, colon, and stomach. Each image measures $1000 \times 1000$ pixels, captured at $40\times$ magnification.

\textbf{\textit{TNBC}} \cite{naylor2018segmentation} dataset comprises nuclei images stained with hematoxylin and eosin, sourced from breast cancer patients. This dataset includes 50 images with a resolution of $512 \times 512$ pixels, captured at $40\times$ magnification. 

\textbf{\textit{CryoNuSeg}} \cite{mahbod2021cryonuseg} dataset contains stained tissues from 10 different organs, providing 30 images of $512 \times 512$ pixels, captured at $40\times$ magnification. The diversity of tissue types offers a comprehensive resource for evaluating the robustness of segmentation methods.

\begin{table*}[!t]
    \centering
    \caption{\textcolor{black}{Comparison with state-of-the-arts on nuclei instance segmentation (Source Domain Generalization).}}
    {\scalebox{0.75}{
    \begin{tabular}{l|c|cccc|cccc|cccc|cccc}
    \toprule
     Datasets & Manual & \multicolumn{4}{c}{$\mathcal{S}_1$} &\multicolumn{4}{|c}{$\mathcal{S}_2$} &\multicolumn{4}{|c}{$\mathcal{S}_3$} & \multicolumn{4}{|c}{$\mathcal{S}_4$}  \\
    \midrule
    Methods & Prompt & AJI & DQ & SQ & PQ & AJI & DQ & SQ & PQ & AJI & DQ & SQ & PQ & AJI & DQ & SQ & PQ\\
    \midrule
    U-Net \cite{ronneberger2015u}       & \multirow{9}{*}{\ding{55}} & 63.49 & 74.81 & 81.72 & 61.45 & 50.27 & 61.75  & 74.74 & 46.23 & 51.34 & 60.73 & 75.32 & 45.88 & 46.43 & 55.90 & 75.80 & 42.38\\
    Mask-RCNN \cite{he2017mask}         &  & 63.32 & 75.03 & 81.05 & 61.44 & 45.32 & 55.98 & 74.16 & 41.61 & 43.84 & 54.79 & 74.71 & 41.35 & 46.72 & 56.16 & \underline{76.25} & 42.87\\
    StarDIST \cite{caicedo2019nucleus}   &  & 63.38 & 74.78 & 80.51 & 60.90 & 54.95 & 68.24 & 74.36 & 50.87 & 45.26 & 56.15 & 76.11 & 42.76 & 46.57 & 53.82 & 74.84 & 40.29\\
    Hover-Net \cite{graham2019hover}     &  & 61.59 & 61.04 & 79.48 & 50.03 & 54.97 & 71.29 & 75.68 & 54.02 & 25.10 & 25.14 & 69.16 & 17.75 & 36.35 & 34.98 & 71.59 & 25.21\\
    TSFD-Net \cite{ilyas2022tsfd}        & & 62.25 & 72.67 & 80.14 & 59.06 & 54.16 & 67.98 & 74.55 & 50.78 & 42.93 & 53.77 & 75.97 & 40.85 & 47.64 & 58.15 & 75.20 & 43.79\\
    CellPose \cite{stringer2021cellpose}   &  & 66.77 & 80.21 & 82.54 & 66.93 & 20.62 & 27.82 & 74.08 & 20.72 & 45.21 & 62.82 & 76.88 & 48.40 & 36.46 & 45.90 & 74.92 & 34.61\\
    CPP-Net \cite{chen2023cpp}           &  & 63.60 & 74.72 & 81.51 & 61.48 & 52.29 & 66.21 & 73.94 & 49.03 & 56.00 & 68.20 & \underline{77.75} & 53.05 & 47.36 & 56.03 & 76.00 & 43.21\\
    CellViT \cite{horst2024cellvit}       &  & 60.51 & 73.92 & \underline{84.10} & 63.25 & 57.91 & 77.35 & \underline{77.19} & 60.54 & 53.16 & 68.72 & 77.65 & 54.83 & 40.87 & 58.13 & 76.16 & 44.38\\ 
    PromptNucSeg \cite{shui2023unleashing}  &           & \underline{74.26} & 85.13 & 82.25 & 70.31 & 64.53 & 79.03 & 76.32 & 60.41 & 61.94 & 72.44 & 77.46  & 56.98 & \underline{52.31} & 62.11 & 73.23 & 45.71\\
    \midrule
    SAM \cite{kirillov2023segment}        &  \multirow{5}{*}{\ding{51}}& 73.32 & 80.21 & 83.08 & 67.16 & 55.51 & 69.70 & 75.24 & 52.51 & 55.16 & 54.44 & 77.74 & 42.45  & 41.66 & 50.71 & 75.71 & 38.77\\
    Med-SA \cite{wu2023medical}           &  & 74.17 & 84.89 & 81.95 & 70.03 & 64.53 & 78.64 & 75.45 & 59.42 & \underline{64.59} & 67.92 & 77.11 & 52.89 & 50.35 & 59.52 & 73.80 & 44.26\\
    SAMed \cite{zhang2023customized}        &    & 71.95 & 82.30 & 80.04 & 66.42 & 62.62 & 75.52 & 75.70 & 57.29 & 63.30 & 65.15 & 75.09 & 49.09 & 49.14 & 56.94 & 73.94 & 42.42\\
    SAMUS \cite{lin2023samus}              &           & 73.70 & \underline{86.97} & 80.88 & \underline{70.72} & \underline{67.98} & \underline{83.09} & 76.87 & \underline{63.93} & 63.34 & \underline{76.56} & 77.23 & \underline{60.84} & 51.25 & \underline{63.53} & 75.79 & \underline{48.35}\\
    SAM-CL \cite{zhong2024convolution}     &           & 73.20 & 86.40 & 80.89 & 70.37 & 63.60 & 78.49 & 74.84 & 58.84 & 60.11 & 67.01 & 77.19 & 51.90 & 50.42 & 61.28 & 74.58 & 45.83\\

    \midrule
    NuSegDG                              &  \ding{51}  & \textbf{77.91} & \textbf{88.88} & \textbf{85.47} & \textbf{76.31} & \textbf{69.81} & \textbf{88.66} & \textbf{77.68} & \textbf{68.88} & \textbf{73.08} & \textbf{85.33} & \textbf{78.15} & \textbf{66.84} & \textbf{53.33} & \textbf{63.64} & \textbf{76.87} & \textbf{49.11}\\
    % NuSegDG (Fine-tuned Auxiliary Network)                         &  \ding{55}  & 78.50 & 89.20 & 86.00 & 77.10 & 70.40 & 89.50 & 78.90 & 69.80 & 73.80 & 85.90 & 78.70 & 67.50 & 54.00 & 64.20 & 77.50 & 50.00\\

    \bottomrule
    \end{tabular}}}
    \label{tab3}
    \vspace{-1.0em}
\end{table*}

\begin{table*}[!t]
    \centering
    \caption{\textcolor{black}{Comparison with state-of-the-arts on nuclei instance segmentation (Target Domain Generalization).}}
    {\scalebox{0.75}{
    \begin{tabular}{l|c|cccc|cccc|cccc|cccc}
    \toprule
     Datasets & Manual & \multicolumn{4}{c}{$\mathcal{T}=\mathcal{S}_1$} &\multicolumn{4}{|c}{$\mathcal{T}=\mathcal{S}_2$} &\multicolumn{4}{|c}{$\mathcal{T}=\mathcal{S}_3$} & \multicolumn{4}{|c}{$\mathcal{T}=\mathcal{S}_4$}  \\
    \midrule
    Methods & Prompt & AJI & DQ & SQ & PQ & AJI & DQ & SQ & PQ & AJI & DQ & SQ & PQ & AJI & DQ & SQ & PQ\\
    \midrule
    U-Net \cite{ronneberger2015u}         & \multirow{9}{*}{\ding{55}} & 6.35  & 4.94  & 10.74 & 3.41  & 31.13 & 28.70 & 66.70 & 19.86 & 14.24 & 16.19 & 48.22 & 10.88  & 42.35 & 47.64 & 72.10 & 34.51\\
    Mask-RCNN \cite{he2017mask}           &  & 7.73  & 7.26  & 12.82 & 5.08  & 30.06 & 30.71 & 66.49 & 21.74 & 11.30 & 14.05 & 48.53 & 9.37   & 43.57 & 48.98 & 71.73 & 35.30\\
    StarDIST \cite{schmidt2018cell}       &  & 8.17  & 6.55  & 11.90 & 4.52  & 35.58 & 39.40 & 70.08 & 28.49 & 18.66 & 24.19 & 56.50 & 16.53  & 43.81 & 49.61 & 71.79 & 35.71\\
    Hover-Net \cite{graham2019hover}       &  & 5.42  & 2.52  & 16.37 & 1.75  & 42.07 & 53.42 & 72.96 & 39.43 & 23.11 & 25.50 & 65.28 & 17.73  & 41.10 & 40.63 & 64.17 & 30.48\\
    TSFD-Net \cite{ilyas2022tsfd}          &  & 6.03  & 5.23  & 11.53 & 3.68  & 41.37 & 49.92 & 71.41 & 35.87 & 12.48 & 17.02 & 51.67 & 11.36  & 37.95 & 45.16 & 71.56 & 32.50\\
    CellPose \cite{stringer2021cellpose}    & & 15.75 & 19.00 & 36.72 & 13.98 & 20.56 & 27.47 & 73.04 & 20.18 & 38.70 & 54.55 & 71.16 & 42.53  & 41.84 & 54.39 & 72.58 & 41.27\\
    CPP-Net \cite{chen2023cpp}            &  & 8.63  & 6.58  & 12.04 & 4.53  & 41.07 & 47.81 & 72.39 & 34.93 & 17.42 & 21.94 & 60.86 & 14.66  & 41.09 & 47.78 & 72.33 & 34.72\\
    CellViT \cite{horst2024cellvit}        &  & 4.39  & 4.95  & 12.00 & 3.66  & 48.44 & \underline{63.31} & 73.85 & \underline{49.17} & 47.30 & 64.00 & 64.76 & 45.53  & 43.94 & 57.23 & 67.57 & 42.36\\ 
    PromptNucSeg \cite{shui2023unleashing} &     & 50.89 & 48.66 & 70.44 & 38.87 & 48.82 & 59.06  & 72.24 & 43.03 & 54.14 & 57.31 & 72.79 & 43.78  & 45.49 & 44.44 & 71.73 & 32.10\\
    \midrule
    SAM \cite{kirillov2023segment}         & \multirow{5}{*}{\ding{51}} & 41.95 & 39.24 & 63.75 & 30.00 & 47.42 & 55.73  & 71.85 & 40.41 & 53.92 & 57.56 & 72.24 & 43.79  & 41.00 & 48.43 & 73.03 & 35.62\\
    Med-SA \cite{wu2023medical}            &  & 57.64 & 55.06 & 74.90 & 44.42 & 50.59 & 62.06  & 74.10 & 46.23 & 56.60 & 63.26 & 72.98 & 47.94  & \underline{51.40} & 61.72 & 73.50 & 45.11\\
    SAMed \cite{zhang2023customized}         &  & 51.15 & 46.98 & 69.71 & 36.92 & 49.47 & 61.00  & 73.58 & 45.10 & 54.59 & 59.69 & 73.20 & 44.14  & 49.34 & 59.31 & 73.33 & 43.65\\
    SAMUS \cite{lin2023samus}               &    & \underline{60.98} & \underline{61.41} & \underline{77.80} & \underline{50.17} & \underline{50.62} & 62.17 & \underline{74.11} & 46.32 & \underline{57.27} & \underline{67.65} & 72.60 & \underline{50.32}  & 51.34 & \underline{62.07} & 73.40 & \underline{45.68}\\
    SAM-CL \cite{zhong2024convolution}    &       & 52.76 & 51.63 & 71.08 & 41.17 & 47.84 & 57.26  & 72.99 & 42.00 & 55.29 & 60.79 & \underline{73.49} & 46.30  & 49.36 & 59.26 & \underline{73.59} & 43.78\\

    \midrule
    NuSegDG                               &   \ding{51}   & \textbf{63.31} & \textbf{72.02} & \textbf{77.99} & \textbf{58.07} & \textbf{58.18} & \textbf{73.19} & \textbf{74.46} & \textbf{54.63} & \textbf{58.30} & \textbf{69.54} & \textbf{73.76} & \textbf{51.61} & \textbf{55.56} & \textbf{63.73} & \textbf{75.78} & \textbf{48.51}\\
    \bottomrule
    \end{tabular}}}
    % \vspace{-2.0em}
    \label{tab4}
\end{table*}

\subsubsection{Implementation Details}
We conduct our experiments on two parallel NVIDIA Tesla P40 GPUs (48GB), utilizing PyTorch 1.13.0, Python 3.10, and CUDA 11.7. We maintain consistent training settings and configurations across all experiments to ensure fairness and reproducibility. For the optimizer, we employ Adam with a batch size of 2 and train models for 100 epochs. The initial learning rate is set to 0.0001 and is adjusted using an exponential decay strategy with a decay factor of 0.98. The loss coefficient $\alpha$ and $\beta$ are set to 0.8 and 0.2 during the training. In our proposed NuSegDG framework, the number of heterogeneous space $N$ is set to 2. All images are resized to $1024 \times 1024$. To save computational costs, the ViT-B \cite{kirillov2023segment} is considered as the image encoder for all SAM-based frameworks. For the TSM-Decoder, we select the pre-trained StarDIST \cite{schmidt2018cell} as our auxiliary neural network to facilitate accurate instance segmentation without manual morphological shape refinement. We utilize the single-point prompt to fine-tune all SAM-based architectures. The point is generated using the \textit{connectedComponents} in OpenCV, which is the centroid of each nucleus instance. \textcolor{black}{For the fluorescence data (which are of single-channel), we replicate the single channel to create an RGB-like input by utilizing the \textit{cvtColor} in OpenCV.}

\subsection{Evaluation Metrics}

In our experiments, we first evaluate the performance of models on the semantic segmentation task using four common metrics: Dice coefficient, mean Intersection over Union (mIoU), F1-score, and Hausdorff Distance (HD). Then, we adopt four extra metrics: Aggregated Jaccard Index (AJI), Detection Quality (DQ), Segmentation Quality (SQ), and Panoptic Quality (PQ), to make comparisons on the instance segmentation task, defined by \cite{horst2024cellvit}:
\textcolor{black}{
\begin{equation}
PQ = \underbrace{\frac{|TP|}{|TP| + \frac{1}{2} |FP| + \frac{1}{2} |FN|}}_{\text{Detection Quality (DQ)}} \times \underbrace{\frac{\sum_{(G, P) \in TP} \text{IoU}(G, P)}{|TP|}}_{ \text{Segmentation Quality (SQ)}},
\end{equation}
where $G$ is the ground truth and $P$ is the prediction segmentation mask. It is important to note that the IoU threshold plays a crucial role in matching predicted instances with ground truth instances, and in our experiments, we have set this threshold to the default value of 0.5 (which is the value that it strikes a balance between being lenient enough to capture true positives and strict enough to penalize poorly placed predictions).} In addition, the best and second-best performance values are highlighted in \textbf{bold} and \underline{underlined}. For each task, we use two different evaluation protocols: domain generalization and adaptability evaluation.

\begin{figure*}[!t]
  \centering
  \includegraphics[width=0.85\linewidth]{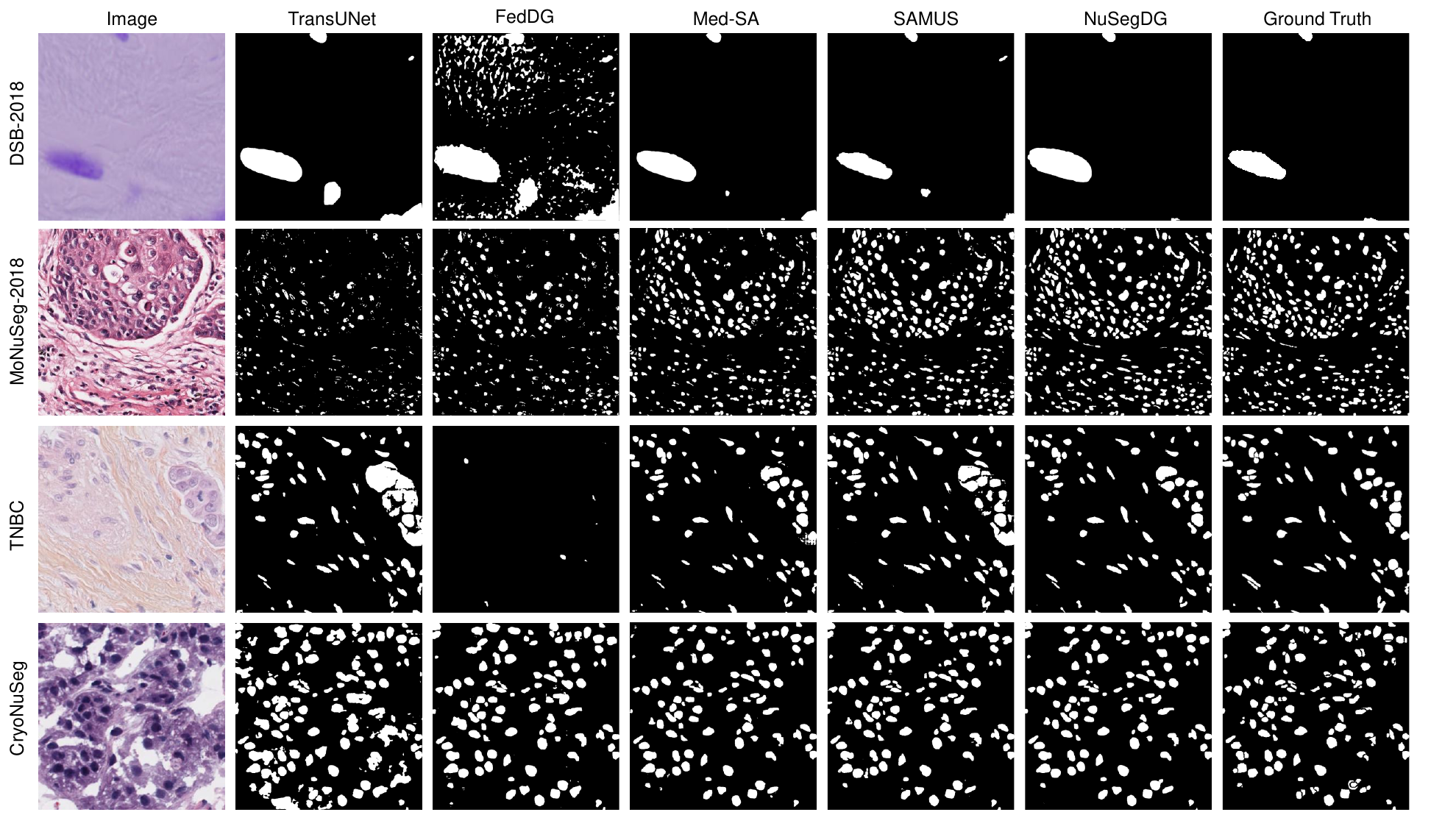}
  \vspace{-0.5cm}
  \caption{Qualitative comparison with state-of-the-art task-specific models and medical SAMs on generalized nuclei semantic segmentation across four target domains: DSB-2018 \cite{caicedo2019nucleus}, MoNuSeg-2018 \cite{kumar2017dataset}, TNBC \cite{naylor2018segmentation} and CryoNuSeg \cite{mahbod2021cryonuseg}.}
  \label{fig:visual}
\end{figure*}

\begin{figure*}[!t]
  \centering
  \includegraphics[width=0.85\linewidth]{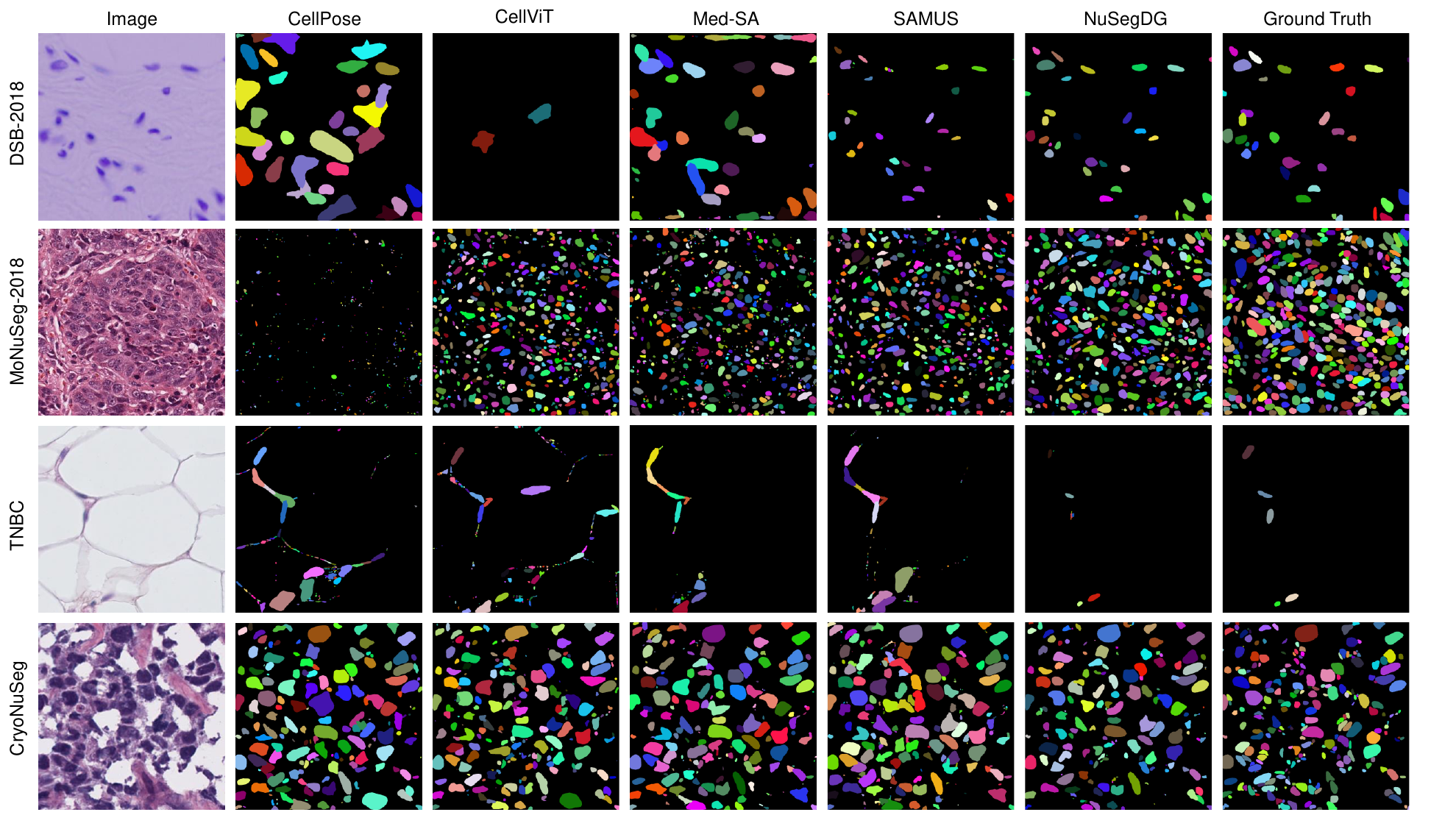}
  \vspace{-0.5cm}
  \caption{Qualitative comparison with state-of-the-art task-specific models and medical SAMs on generalized nuclei instance segmentation across four target domains: DSB-2018 \cite{caicedo2019nucleus}, MoNuSeg-2018 \cite{kumar2017dataset}, TNBC \cite{naylor2018segmentation} and CryoNuSeg \cite{mahbod2021cryonuseg}.}
  \label{fig:visual2}
  \vspace{-1.0em}
\end{figure*}

\subsubsection{Source Domain Generalization Evaluation}
In this protocol, we perform a fully supervised learning where all four datasets (i.e., $\mathcal{S}_1, \mathcal{S}_2, \mathcal{S}_3,\mathcal{S}_4$) are considered as seen domains. We randomly divide all datasets into three sets: training, validation, and testing, in the conventional ratio of 8:1:1. The model is evaluated on the testing set of each dataset individually. The reason for conducting this protocol is to assess the adaptability of each model across different domains. Moreover, we display the performance gap between our domain generalization approach and traditional fully supervised methods. This comparison further demonstrates the effectiveness of NuSegDG on domain generalization tasks.

\subsubsection{Target Domain Generalization Evaluation}
We employ a standard leave-one-domain-out strategy \cite{zhou2022domain} to conduct the domain generalization evaluation. Specifically, the model is trained on a training set $\mathcal{S}$ of $K-1$ source domains, where each source domain represents a different data distribution, and then evaluated on the remaining unseen target domains $\mathcal{T}$, e.g, $\mathcal{S}=\{\mathcal{S}_1, \mathcal{S}_2, \mathcal{S}_3\}$, $\mathcal{T}=\mathcal{S}_4$.

\begin{table}[!t]
    \centering
    \small
    \caption{Ablation study of NuSegDG in domain-generalized Nuclei Instance Segmentation: $\mathcal{S} \rightarrow \mathcal{T}$. $M_1$: HS-Adapter. $M_2$: GKP-Encoder. $M_3$: TSM-Decoder.}
    {\scalebox{0.81}{
    \begin{tabular}{cccc|cccc}
    \toprule
    Row & $M_1$ & $M_2$ & $M_3$ & AJI (Avg.) & DQ (Avg.) & SQ (Avg.) & PQ (Avg.)\\
    \midrule
    1 & &  &  & 48.69 & 53.78 & 70.11 & 39.90\\ % 86.43 4.15
    2 & \checkmark &  &  & 53.86 & 60.14 & 72.77 & 46.51\\ % 24.93 0.02 3.51
    3 & & \checkmark &  & 51.48 & 56.66 & 71.36 & 43.07\\
    4 & & & \checkmark & 50.70 & 55.81 & 71.05 & 41.48\\
    5 & \checkmark & \checkmark & & 56.29 & 66.23 & 74.21 & 50.38\\
    6 & \checkmark & & \checkmark & 54.85 & 63.93 & 73.17 & 48.26\\
    7 & & \checkmark & \checkmark & 53.15 & 58.76 & 71.96 & 46.14\\
    8 & \checkmark & \checkmark & \checkmark & \textbf{58.84} & \textbf{69.62} & \textbf{75.50} &\textbf{53.21}\\
    \bottomrule
    \end{tabular}}}
    \label{tab:7}
    % \vspace{-2.0em}
\end{table}

\subsection{Comparison on Nuclei Semantic Segmentation}

To comprehensively assess our NuSegDG, Table \ref{tab1} provides generalization evaluation results on source domains. We observe that previous U-shape architectures show remarkable performance gains in the seen domain but are inferior to PEFT SAMs. Our NuSegDG achieves superior performance on these four datasets, with the best mIoU of 87.46\%, 76.06\%, 78.93\% and 73.44\%, respectively. On the other hand, the domain-generalized NuSegDG in Table \ref{tab2} demonstrates competitive performance on $\mathcal{S}_2$, $\mathcal{S}_3$ and $\mathcal{S}_4$ domains compared to fully-supervised U-shape and SAM-based architectures in Table \ref{tab1}. We provide the visualization results in Fig. \ref{fig:visual}.

Moreover, we compare it with state-of-the-art frameworks on nuclei semantic segmentation. As illustrated in Table \ref{tab2}, in the target domain generalization evaluation, TransUNet \cite{chen2024transunet} achieves leading results among previous U-shape segmentation algorithms due to its large model capacity. Benefiting from pre-training on the large-scale dataset, PEFT SAMs \cite{wu2023medical, zhang2023customized, lin2023samus, zhong2024convolution, na2024segment,gu2024lesam, cheng2024unleashing} display better performance than these task-specific models. In contrast, our NuSegDG surpasses the second-best SAMUS by a significant mIoU increase of 2.62\%, 3.07\%, 2.55\%, and 4.51\% on these four target domains, respectively. Compared to the prompt-free SAMs, NuSegDG presents a mIoU rise of 5.01\% to 12.57\%. Consequently, these comparisons validate the superiority of our NuSegDG on domain-generalized nuclei semantic segmentation tasks in diverse nuclei domains.

\begin{table}[!t]
    \centering
    \small
    \caption{Comparison of inference time with the vanilla Point Prompt mode.}
    {\scalebox{0.9}{
    \begin{tabular}{l|cccc}
    \toprule
    Datasets & $\mathcal{S}_1$ &$\mathcal{S}_2$ & $\mathcal{S}_3$ & $\mathcal{S}_4$\\
    \midrule
    Nuclei (Avg.) & 30 & 510 & 99 & 157\\   
    \midrule
    Vanilla Point Prompt & 11.19s & 61.49s & 12.64s & 19.92s\\ % 86.43 4.15
    $+$ Auxiliary Neural Network & \textbf{10.31s} & \textbf{11.62s} & \textbf{10.55s} & \textbf{10.86s}\\ % 24.93 0.02 3.51
    \bottomrule
    \end{tabular}}}
    \label{tab:6}
    % \vspace{-2.0em}
\end{table}

\begin{table*}[!t]
    \centering
    \caption{\textcolor{black}{Comparison with the original point prompt mode on nuclei semantic segmentation (Source Domain Generalization).}}
    {\scalebox{0.82}{
    \begin{tabular}{c|cccc|cccc|cccc|cccc}
    \toprule
     Datasets & \multicolumn{4}{c}{ $\mathcal{S}_1$} &\multicolumn{4}{|c}{$\mathcal{S}_2$} &\multicolumn{4}{|c}{$\mathcal{S}_3$} & \multicolumn{4}{|c}{$\mathcal{S}_4$}  \\
    % \cmidrule(lr){1-7}
    \midrule
    Prompt Types &  Dice & mIoU & F1 & HD & Dice & mIoU & F1 & HD &  Dice & mIoU & F1 & HD & Dice & mIoU & F1 & HD\\
    \midrule

    1 point  & 92.09 & 85.62 & 92.20 & 58.29 & 84.45 & 73.17 & 84.44 & 59.96 & 85.53 & 74.76 & 85.48 & 85.28 & 83.08 & 71.32 & 83.43 & 90.49\\
    3 points   & 92.66 & 86.78 & 92.71 & 60.49 & 84.38 & 73.12 & 84.41 & 59.64 & 86.77 & 76.67 & 86.62 & 101.42 & 82.85 & 71.04 & 82.98 & 91.97\\
    5 points & 92.93 & 87.08 & 92.95 & 41.74 & 84.40 & 73.11 & 84.43 & 39.01 & 86.99 & 77.03 & 86.92 & 51.73 & 84.09 & 72.86 & 84.18 & 86.73\\
    7 points  & 93.08 & 87.26 & 93.11 & 53.27 & 85.06 & 74.07 & 85.09 & 55.81 & 87.57 & 77.94 & 87.50 & 59.45 & 84.22 & 72.93 & 84.30 & 83.39\\
    \midrule
    1 point + GK & \textbf{93.17} & \textbf{87.46} & \textbf{93.35} & \textbf{33.18} & \textbf{86.37} & \textbf{76.06} & \textbf{86.40} & \textbf{44.44} & \textbf{88.20} & \textbf{78.93} & \textbf{87.03} & \textbf{49.69} & \textbf{84.59} & \textbf{73.44} & \textbf{84.75} & \textbf{64.61}\\
      % NuSegDG &\textbf{87.06} & \textbf{92.90} & \textbf{93.13} & \textbf{93.13} &\textbf{75.28} & \textbf{85.85} & \textbf{84.99} & \textbf{86.86} &\textbf{77.01} & \textbf{86.98} & \textbf{89.46} & \textbf{84.72}\\
    \bottomrule
    \end{tabular}}}
    % \vspace{-1.0em}
    \label{tab8}
\end{table*}

\begin{table*}[!h]
    \centering

    \caption{\textcolor{black}{Comparison of pretrained and fine-tuned auxiliary network on nuclei instance segmentation (Source Domain Generalization).}}
    {\scalebox{0.76}{
    \begin{tabular}{l|c|cccc|cccc|cccc|cccc}
    \toprule
     Datasets &  & \multicolumn{4}{c}{$\mathcal{S}_1$} &\multicolumn{4}{|c}{$\mathcal{S}_2$} &\multicolumn{4}{|c}{$\mathcal{S}_3$} & \multicolumn{4}{|c}{$\mathcal{S}_4$}  \\
    \midrule
    Methods & Tuned & AJI & DQ & SQ & PQ & AJI & DQ & SQ & PQ & AJI & DQ & SQ & PQ & AJI & DQ & SQ & PQ\\
    \midrule
    \multirow{2}{*}{Auxiliary Network}        & \ding{51} & 77.91 & 88.88 & 85.47 & 76.31  & 69.81 & 88.66 & 77.68 & 68.88 & 73.08 & 85.33 & 78.15 & 66.84   & 53.33 & 63.64 & 76.87 & 49.11 \\
               & \ding{55} & 78.50 & 89.20 & 86.00 & 77.10  & 70.40 & 89.50 & 78.90 & 69.80 & 73.80 & 85.90 & 78.70 & 67.50  & 54.00 & 64.20 & 77.50 & 50.00\\
    
    \bottomrule
    \end{tabular}}}
    % \vspace{-2.0em}
    \label{tab8}
\end{table*}

\subsection{Comparison on Nuclei Instance Segmentation}

To further evaluate our NuSegDG in nuclei instance segmentation tasks, we provide the source domain generalization comparison result in Table \ref{tab3}. It is demonstrated that PEFT SAMs outperform morphological post-processing algorithms in four nuclei datasets. For example, PromptNucSeg \cite{shui2023unleashing} has a 6.62\% AJI increase over CellViT \cite{horst2024cellvit} on the $\mathcal{S}_2$ domain. In contrast, our NuSegDG framework achieves the best AJI of 77.91\%, 69.81\%, 73.08\%, and 53.33\%, respectively, on the four datasets, and performs better than the state-of-the-art methods in the other three evaluation metrics. 

Furthermore, we perform the comparison with advanced nuclei instance segmentation frameworks on four different nuclei domains. Firstly, Table \ref{tab4} presents experimental results under the target domain generalization evaluation. It is revealed that previous morphological post-processing algorithms \cite{graham2019hover, ilyas2022tsfd, stringer2021cellpose, chen2023cpp, horst2024cellvit} show poor generalization capabilities on the $\mathcal{S}_1$ domain. On the contrary, SAMUS \cite{lin2023samus} performs better than these methods by achieving a remarkable PQ of 50.17\%, 46.32\%, 50.32\% and 45.68\% on four domains, respectively. Our NuSegDG outperforms it with a significant PQ increase of 7.90\%, 8.31\%, 1.29\%, and 2.83\%, respectively. The quantitative comparison is presented in Fig. \ref{fig:visual2}. As a result, these results reveal a significant performance advantage of our NuSegDG over current medical foundation models and task-specific architectures on domain-generalized nuclei semantic and instance segmentation tasks.

\subsection{Ablation Study}

To investigate the effectiveness of the individual components within the NuSegDG framework, we conduct an ablation study on domain-generalized nuclei instance segmentation, as summarized in Table \ref{tab:7}. This study sequentially enables or disables the HS-Adapter $M_1$, GKP-Encoder $M_2$, and TSM-Decoder $M_3$ to evaluate their impact on the performance of the average AJI, DQ, SQ, and PQ metrics. Firstly, we consider the standard fine-tuned SAM ($1^{st}$ row) as the ablation baseline. By respectively embedding the HS-Adapter ($2^{nd}$ row), GKP-Encoder ($3^{rd}$ row) and TSM-Decoder
($4^{th}$ row), the performance is raised with the average AJI of 5.17\%, 2.79\%, 2.07\%, and the average PQ of 6.61\%, 3.17\%, 1.58\%. When we combine HS-Adapter with GKP-Encoder ($5^{th}$ row), the performance of the model is further improved, with the average AJI of 56.29 and PQ of 50.38\% on the four domains. This result proves that these two modules can promote the domain generalization capability in nuclei instance segmentation. By comparing $6^{th}$ and $7^{th}$ rows with $2^{nd}$ and $3^{rd}$ rows, the TSM-Decoder demonstrates significant performance gains while eliminating the demand for manual morphological refinement. Finally, our NuSegDG framework ($8^{th}$ row) integrates all three modules and achieves the best performance on all metrics, with an average AJI of 58.84\%, an average DQ of 69.62\%, an average SQ of 75.50\%, and an average PQ of 53.21\%. This full configuration significantly outperforms the others, emphasizing the synergistic benefits of incorporating all modules. This result highlights the importance of each component in enhancing the generalization capability of NuSegDG across different nuclei image domains.

Moreover, nuclei semantic and instance segmentation tasks meet two different requirements for clinical applications, such as disease area calculation and nuclei counting. Based on experimental results, we can observe that existing automatic SAMs show competitive generalization performance on source domains but are inferior to the SAMs with point prompts on target domains. On the other hand, although the vanilla point prompt performs better in domain generalization, it predicts cell instances one by one which is time-consuming for dense cell maps (e.g., whole slide imaging), as demonstrated in Table \ref{tab:6}. By introducing our auxiliary neural network, our NuSegDG framework achieves remarkable generalization-efficiency trade-offs.

\textcolor{black}{In addition to the ablation study on the individual components of NuSegDG, we also conduct a supplementary experiment to compare the traditional point prompt approach used in the original SAM with our proposed density map prompt. As shown in Table 7, the results indicate that while increasing the number of point prompts from one to seven can lead to slight improvements in segmentation performance, the use of the density map prompt consistently achieves the best results across all metrics. For instance, the density map prompt yields a Dice score of 93.17\%, an mIoU of 87.46\%, and an F1 score of 93.35\%, while substantially reducing the Hausdorff Distance (HD) to 33.18, which is significantly lower than those obtained by any point prompt configuration. This experiment demonstrates that the density map prompt not only captures richer semantic and positional cues compared to single or multiple point prompts but also substantially improves segmentation accuracy and boundary adherence. Consequently, this reinforces the effectiveness of our Gaussian-Kernel Prompt Encoder in providing robust and efficient guidance for segmentation decoding, so that it contributes to the overall domain generalization capability of the NuSegDG framework.}

\subsection{Analysis of Hyper-Parameters}

\begin{figure}[!t]
  \centering
  \includegraphics[width=1\linewidth]{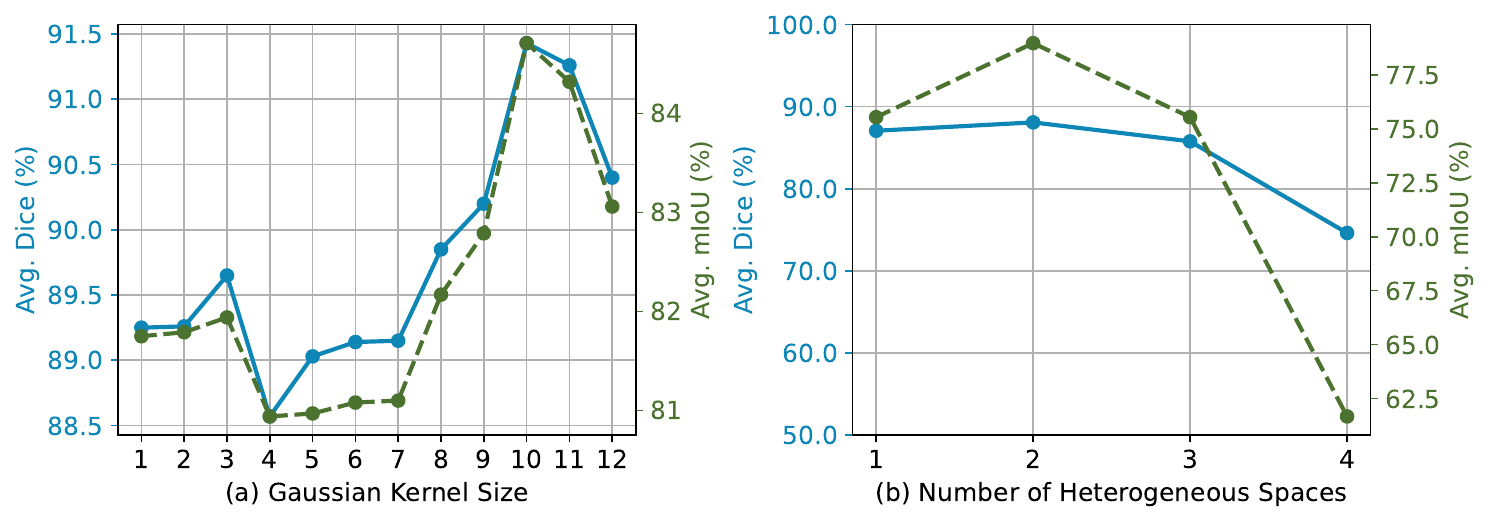}
  \caption{Hyper-parameter analysis of kernel size in GKP-Encoder (a) and number of learnable parameters in HS-Adapter (b).}
  \label{fig:diss}
  % \vspace{-2.0em}
\end{figure}

In this section, we perform a comprehensive hyper-parameters analysis of our NuSegDG model. As reported in Section \ref{m1} and \ref{m2}, NuSegDG contains two hyper-parameters, including the Gaussian kernel size $r$ in GKP-Encoder and the number of heterogeneous space $N$ in HS-Adapter. For the kernel size, we perform a grid search under the fully-supervised learning to select an optimal configuration. Fig. \ref{fig:diss}a shows the average Dice and mIoU of NuSegDG on the four nuclei domains with different kernel sizes. It is indicated that the NuSegDG with $r=10$ demonstrates the best performance due to the sufficient semantic information prompts. However, excessive kernel size may generate false positive errors, which cannot offer additional benefits. For the number of heterogeneous space, we provide the result of grid search in Fig. \ref{fig:diss}b. We observe that the NuSegDG with $N=2$ obtains the best performance. Setting more heterogeneous space significantly increases the computational complexity of NuSegDG, which is not suitable for limited training samples in nuclei domains. These experimental results prove the importance of tuning these hyper-parameters to improve the efficiency of our NuSegDG framework in learning domain-specific knowledge. \textcolor{black}{In addition, compared to existing fully fine-tuned methods (e.g., MedSAM \cite{ma2024segment}, SAMMI \cite{huang2024segment}), NuSegDG only requires a small number of trainable parameters, significantly reducing computational costs. Moreover, the standard SAM \cite{kirillov2023segment} and other variants (e.g., LeSAM \cite{gu2024lesam}, SAM-CL \cite{zhong2024convolution}, Med-SA \cite{wu2023medical}) need multiple point prompts to generate accurate segmentation results. On the contrary, NuSegDG can realize better performance with a single-point prompt, which is more effortless and helps reducing annotation time and fatigue. Further, we fine-tune the Auxiliary Network (AN) on four nuclei instance segmentation datasets. The result has been presented in Table \ref{tab8}. Our findings indicate that fine-tuning AN yields a modest improvement in performance across several metrics (e.g., AJI, DQ, SQ, and PQ) compared to using the pretrained weights. This suggests that while fine-tuning can help further optimize the performance, the pretrained model already captures critical features required for generating accurate instance maps. Consequently, our initial design choice of using pretrained weights remains justified in terms of computational efficiency, with the fine-tuned version providing an upper-bound performance reference.} 

\section{Conclusion}
In this paper, we have proposed NuSegDG for domain-generalized nuclei image segmentation. Specifically, the HS-Adapter has been introduced to adapt the feature representation of SAM from natural to different nuclei images by heterogeneous space integration. Then, the GKP-Encoder has been devised to produce high-quality density maps, driven by the single-point prompt, with sufficient semantic information for guiding segmentation predictions. Finally, the TSM-Decoder has achieved the automatic conversion between the semantic masks and instance maps without demand for labor-intensive morphological post-processing methods. Extensive experimental results have demonstrated that NuSegDG has outperformed the existing nuclei-specific and SAM-based segmentation methods in domain-generalized nuclei image segmentation and displayed superior adaptability across different nuclei domains. \textcolor{black}{The proposed NuSegDG presents a potential nuclei annotation tool for improving the efficiency of data labeling, and its accurate delineation of nuclei can aid in tumor detection, grading, and diagnostic assessments.}

\section*{Declaration of competing interest}
The authors declare that they have no known competing financial interests or personal relationships that could have
appeared to influence the work reported in this paper.
\section*{Acknowledgements}
This work is partially supported by the Yongjiang Technology Innovation Project (2022A-097-G), Zhejiang Department of Transportation General Research and Development Project (2024039), and National Natural Science Foundation of China grant (UNNC ID: B0166).

\bibliographystyle{unsrt}
\bibliography{mybib}
	
\end{document}